\documentclass[preprint,review,12pt]{elsarticle}

\usepackage{amssymb}
\usepackage{upgreek}

\usepackage{lineno}

\journal{Nuclear Instruments and Methods A}

\begin{document}

\begin{frontmatter}



 \title{Gain Recovery in Heavily Irradiated Low Gain Avalanche Detectors by High Temperature Annealing \tnoteref{t1}}
 \tnotetext[t1]{The work was partly done in the framework of the RD50 collaboration.}

 \author[ijs]{I.~Mandi\' c\corref{cor1}}
 \ead{igor.mandic@ijs.si}

 \author[ijs]{V. ~Cindro}
 \author[ijs]{A. ~Gori\v sek}
 \author[ijs]{B. ~Hiti}
 \author[ijs]{A. ~Howard}
  \author[ijs]{\v Z. ~Kljun}
 \author[ijs]{G. ~Kramberger}
 \author[ijs]{M. ~Ma\v cek Kr\v zmanc}
 \author[ijs,fmf]{M. ~Miku\v z}
\author[ijs]{B. ~Novak}

\cortext[cor1]{Corresponding author}

\address[ijs]{Jo\v zef Stefan Institute, Jamova 39, Ljubljana, Slovenia}
\address[fmf]{University of Ljubljana, Faculty of Mathematics and Physics, Jadranska 19, Ljubljana, Slovenia}


\begin{abstract}
Studies of annealing at temperatures up to 450$^\circ$C with Low Gin Avalnche Detectors (LGAD) irradiated with neutrons are described. It was found that the performance of lgads irradiated with 1.5e15 n/cm$^2$ was already improved at 5 minutes of annealing at 250$^\circ$C. Isochronal annealing for 30 minutes in 50$^\circ$C steps between 300$^\circ$C  and 450$^\circ$C showed that the largest beneficial effect of annealing is at around 350$^\circ$C. Another set of devices was annealed for 60 minutes at 350$^\circ$C and this annealing significantly increased $V_{\mathrm{gl}}$. The effect is equivalent to reducing the effective acceptor removal constant by a factor of $\sim$ 4. Increase of  $V_{\mathrm{gl}}$ is the consequence of increased effective space charge in the gain layer caused by formation of electrically active defects or re-activation of interstitial boron atoms. 
\end{abstract}


\end{frontmatter}


\section{Introduction}
\label{intro}

In Low Gain Avalanche Detector (LGAD) charge multiplication (gain) is achieved by introducing a thin highly doped ($\sim 10^{16}$ cm$^{-3}$) layer of p-type silicon within high resistivity p-type bulk under the charge collecting n-type electrode \cite{lgad}. The doping of the gain layer is tailored so that an electric field strength of a  few 10 V/$\mu $m is established in the layer when it is depleted. Such a field leads to charge multiplication by impact ionisation of electrons, with a gain factor of up to 100. At the same time the electric field strength is too low for significant multiplication of holes which would lead to breakdown. Moderate gain for electrons enables efficient detection of MIPs with very thin silicon detectors (50 $\mu$m or less) and provides rapid signals enabling good signal-to-noise ratio and short dead time. This results in efficient charged particle detection with good time resolution even in a high radiation and high rate hadron collider environment. Therefore, LGAD has been chosen as the technology for timing detectors in both ATLAS and CMS for the upgrade to HL-LHC \cite{Hartmut, HGTD_TDR, TIMING_TDR_CMS}. Timing resolution for MIP passage of few tens of ps can be reached with LGADs. Such an excellent timing resolution can also be exploited in many other applications.   

The main drawback for usage of LGADs in hadron colliders is their sensitivity to displacement damage caused by energetic hadrons. The problem is the loss of gain due to radiation induced acceptor removal in the gain layer \cite{LgadKrambi}. The dominant process leading to gain degradation is interaction of substitutional boron with an interstitial resulting in electrically inactive interstitial boron ($ I + B_S \rightarrow B_I$) which can further form different defect complexes. 
Smaller space charge concentration leads to smaller electric field in the gain layer and eventually to loss of gain. This limits the operation of LGADs in timing detectors at HL-LHC to a maximal fluence of 3$\cdot 10^{15}$ n$_{\mathrm{eq}}$/cm$^2$ (1 MeV neutron equivalent) \cite{HGTD_TDR}.

It is well known that the effects of radiation damage change with time and that the process is temperature dependent. The effect of annealing at 60$^\circ$C on gain, up to over 20000 minutes, has been studied and only moderate effects on LGAD performance were observed  \cite{lgad_annealing}.
However, annealing at higher temperatures might have more significant effects and in this work we report about measurements with irradiated LGADs annealed at temperatures up to 450$^\circ$C.

Studies of the possibility to recover radiation damage in silicon detectors by annealing at elevated temperatures have been done before. In \cite{DRIVE} authors discuss the feasibility of Detector Recovery/Improvement Via Elevated-Temperature-Annealing (DRIVE) approach. It was found that annealing at 450$^\circ$C leads to sign inversion of the effective space charge in p-type silicon and to lower full depletion voltage. Both effects may result in improved charged particle detection performance. The effects were explained with introduction of thermal donors in oxygen-rich MCZ material. Thermal donors compensate radiation induced negative space charge as well as the negative space charge introduced by the reverse annealing \cite{ROSE}.  

It may be expected that high temperature annealing will be beneficial for performance of LGADs because its degradation is caused by the drop of negative space charge concentration in the gain layer. This could be compensated by reverse annealing processes i.e. activation of effective acceptors on the long time scale. Moreover, it was shown in \cite{LgadKrambi} that higher $N_{\mathrm{eff}}$ in the bulk of LGAD may also lead to lower operational voltage needed to establish sufficient gain. 

Additional motivation for studies at high temperature was the need of future measurements with ASIC+LGAD assemblies with irradiated sensors. These are needed for the development of ATLAS High Granularity Timing Detector (HGTD)  \cite{HGTD_TDR}. In the assemblies LGAD is bump bonded to the ASIC and bonds contain silver (Ag) which forms long lived isotopes after irradiation with neutrons. Studies would be greatly simplified if LGAD could be irradiated separately and assembled with the ASIC after irradiation. But the bonding process requires heating of the assembly to 250$^\circ$C for a few minutes which might affect the performance of irradiated LGAD. To check the effect a study of 5 minutes annealing at 250$^\circ$C was performed and the results are included in this paper.

\section{Samples, irradiation}
\label{samples}

\begin{figure}[!hbt]
\centering
 \includegraphics[width=0.7\textwidth]{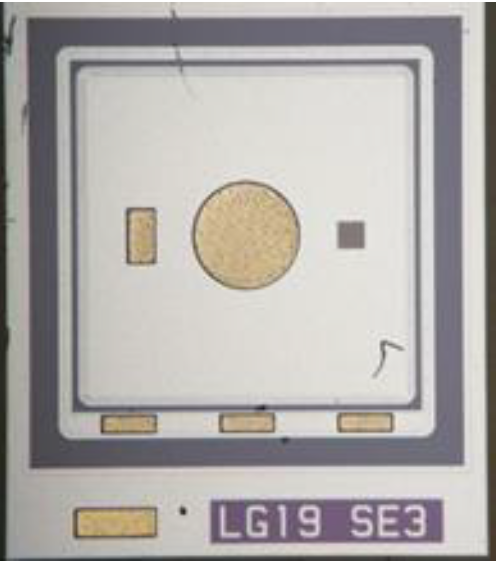}
 \caption{\label{fig1} Photo of HPK device used in this work. The sensitive area is 1.3 $\times$ 1.3 mm$^2$, active thickness is 50 $\mu$m.}
 
\end{figure}

Single pad  LGADs with a sensitive area of 1.3 $\times$ 1.3 mm$^2$, an active thickness of 50 $\mu$m and total thickness of the device 200 $\mu$m, produced by HPK \cite{HPK} were used for this study. An example is shown in the photo in Fig. \ref{fig1}. Voltages required for depletion of the gain layer $V_{\mathrm{gl}}$  were between 40 and 55 V before irradiation while full depletion voltages $V_{\mathrm{fd}} $ were between 45 to 65 V. Samples were irradiated with neutrons in the TRIGA reactor in Ljubljana \cite{Reactor1, Reactor2}. Table \ref{tab1} summarizes the properties of the devices and the 1 MeV neutron equivalent fluences to which they were exposed.

 \begin{table}[htbp]
\centering
\caption{\label{tab1} Properties of investigated samples. active thickness 50 $\mu$m. V$_\mathrm{gl}$ is the depletion voltage of gain layer in LGAD.}
\smallskip
\begin{tabular}{l| c c c c c}
  \hline
Device type & V$_{\mathrm{gl}}$ (V) & Gain layer & V$_{\mathrm{fd}}$ (V) & $\Phi_{\mathrm{eq}}$ (10$^{15}$ n/cm$^2$) \\
& & depth  ($\mu$m)  & & \\
\hline
HPK-P1-T3.1 & 41 & 1.6 & 49 & 0.8, 3, 6 \\
HPK-P1-T3.2 & 55.5 & 2.4 & 64 & 1.5, 4, 6 \\
HPK-P2-W28 & 54.5 & 2.4 & 60 &  1.5 \\ 

\hline
\end{tabular}
\end{table}


\section{Experimental techniques}
\label{techniques}

Samples were characterised by capacitance-voltage (CV) and current-voltage (IV) measurements on the probe station. They were placed on a chuck with temperature stabilized to (20 $\pm$ 0.2)$^\circ$C. Keithley 6517A Electrometer was used as the high voltage source and the current meter. Capacitances were measured with HP 4263B LCR meter at 10 kHz.

Timing properties were investigated with signals generated by MIP-like electrons from a $^{90}$Sr source. Samples were wire-bonded to a dedicated electronics board designed by UCSC \cite{ucsc} and cooled to -30$^\circ$C. Signals generated by passage of electrons were processed with a fast trans-impedance amplifier followed by a commercial amplifier (Particulars AM-02B, 35 dB, bandwidth $>$ 3 GHz). The signals from the amplification chain were digitised by a 20 GS/s digitising oscilloscope with 2.5 GHz bandwidth. For measurement of timing resolution, signals from the investigated sensors were compared with signals from the reference detector with known timing resolution. Timing resolution was determined from the spread of the difference in time of arrival between the investigated and the reference detector. More details about the system and measurement method can be found in \cite{lgad_annealing}.

Heat treatment of samples was made in air or Argon atmosphere, depending on the equipment used. Heating in air was performed in a laboratory drying oven (SP-45 C, Kambič, Slovenia). The heat treatments in Argon atmosphere were performed in the furnace for Simultaneous thermal analysis (STA) Jupiter 449 (Netzsch, Selb, Germany). No study of the effect of different atmospheres was done in this work. 


\section{Measurements}
\label{measurements}

\subsection{Annealing for 5 minutes at 250$^\circ$C}

\begin{figure}[!hbt]
\centering
\begin{tabular}{c c } 
 \includegraphics[width=0.5\textwidth]{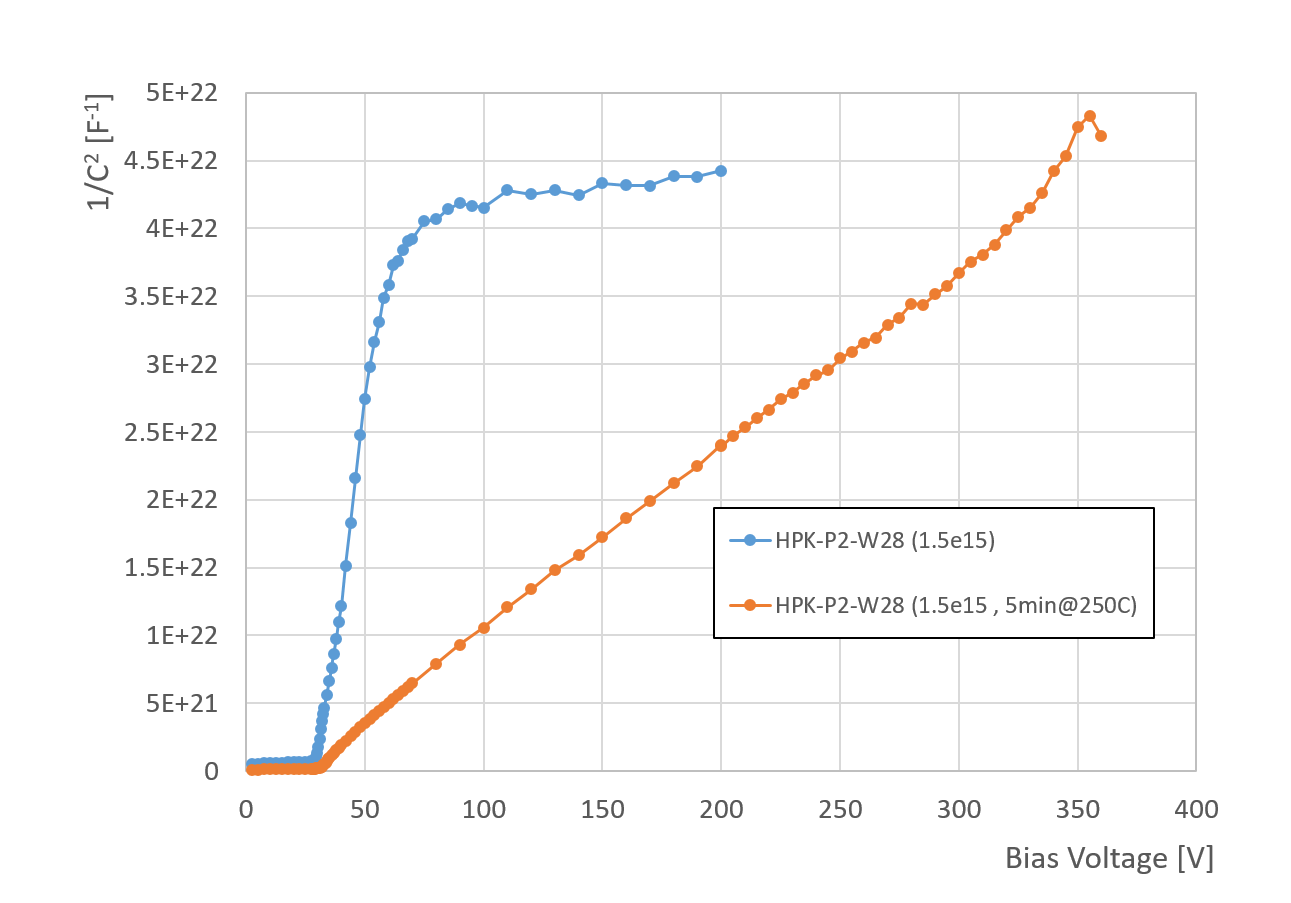} &  \includegraphics[width=0.5\textwidth]{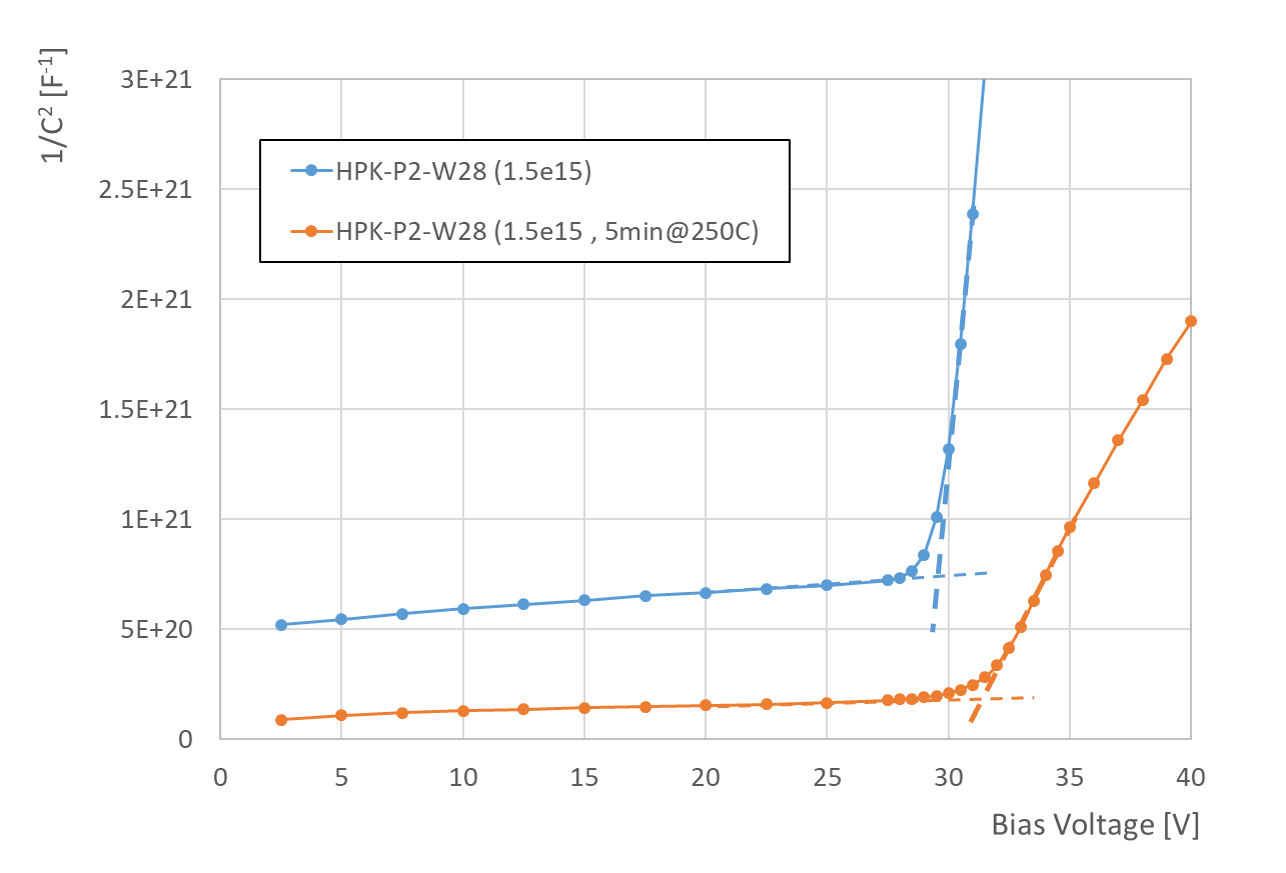} \\
 a) & b)  \\
 \includegraphics[width=0.5\textwidth]{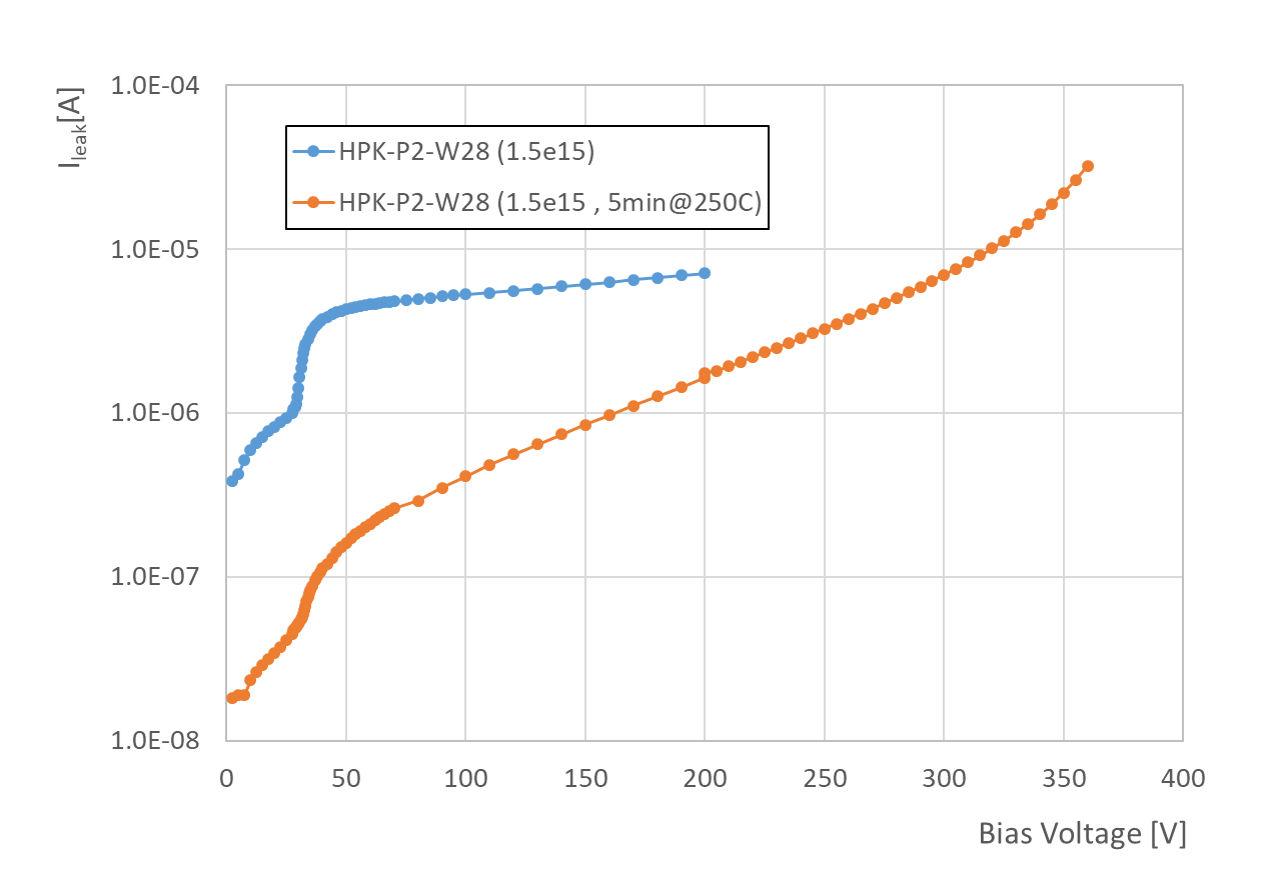} &  \\
 c) &  
 \end{tabular}
 \caption{\label{CVIV-5min} CV/IV measurements (blue) before and after annealing for 5 minutes at 250$^\circ$C (orange) with device irradiated to 1.5$\cdot10^{15}$ n$_\mathrm{eq}$/cm$^2$. Figure a) shows C-V measurements and b) shows zoom to low bias voltages. Dashed lines are fits to straight parts of the curve. The intersection of these lines defines the gain layer depletion voltage $V_{\mathrm{gl}}$. Figure c) shows I-V measurements before and after annealing. Measurements were taken at 20$^\circ$C. }
\end{figure}

Figure \ref{CVIV-5min} shows results of CV-IV measurements with HPK-P2-W28 annealed for 5 minutes at 250$^\circ$C.
\begin{figure}[!hbt]
\centering
\begin{tabular}{c c } 
 \includegraphics[width=0.5\textwidth]{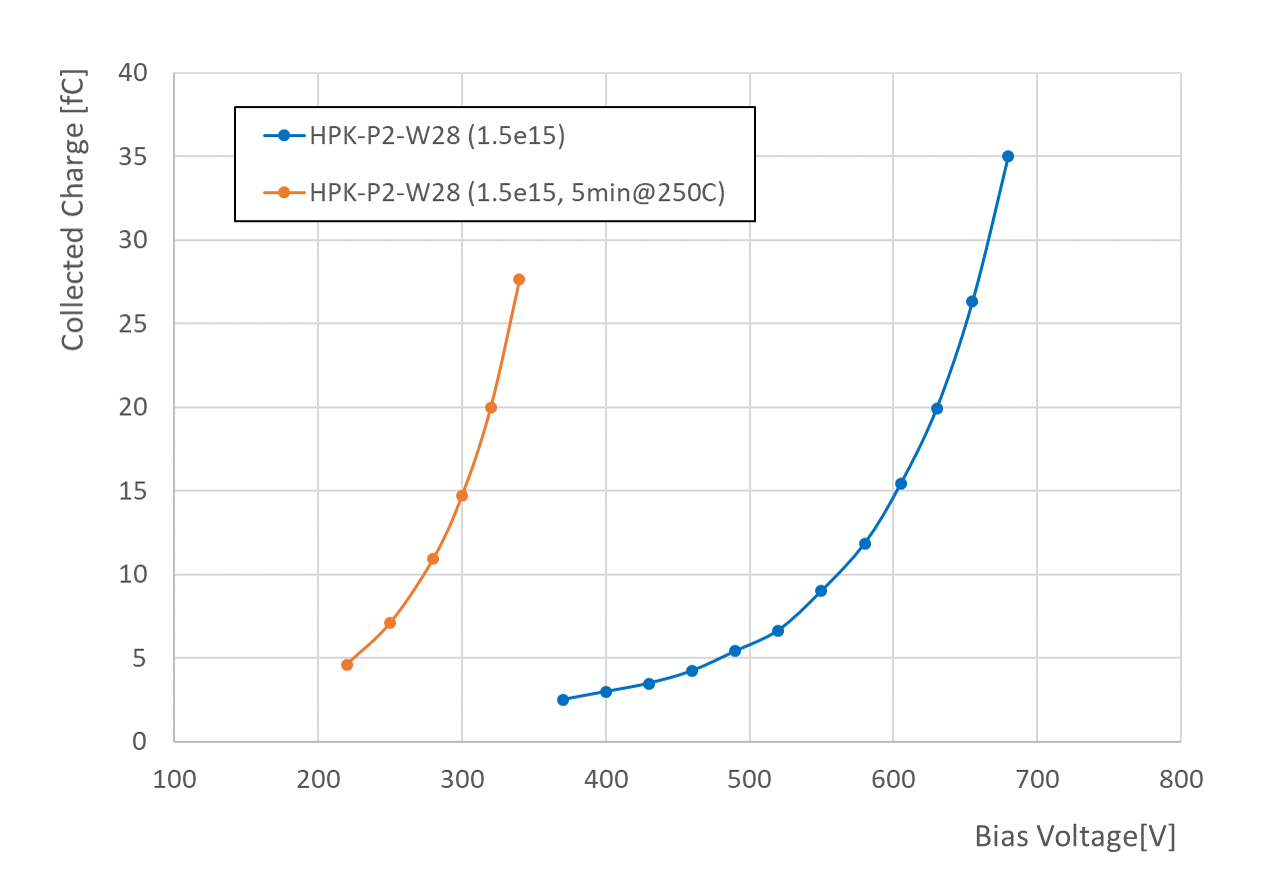} & \includegraphics[width=0.5\textwidth]{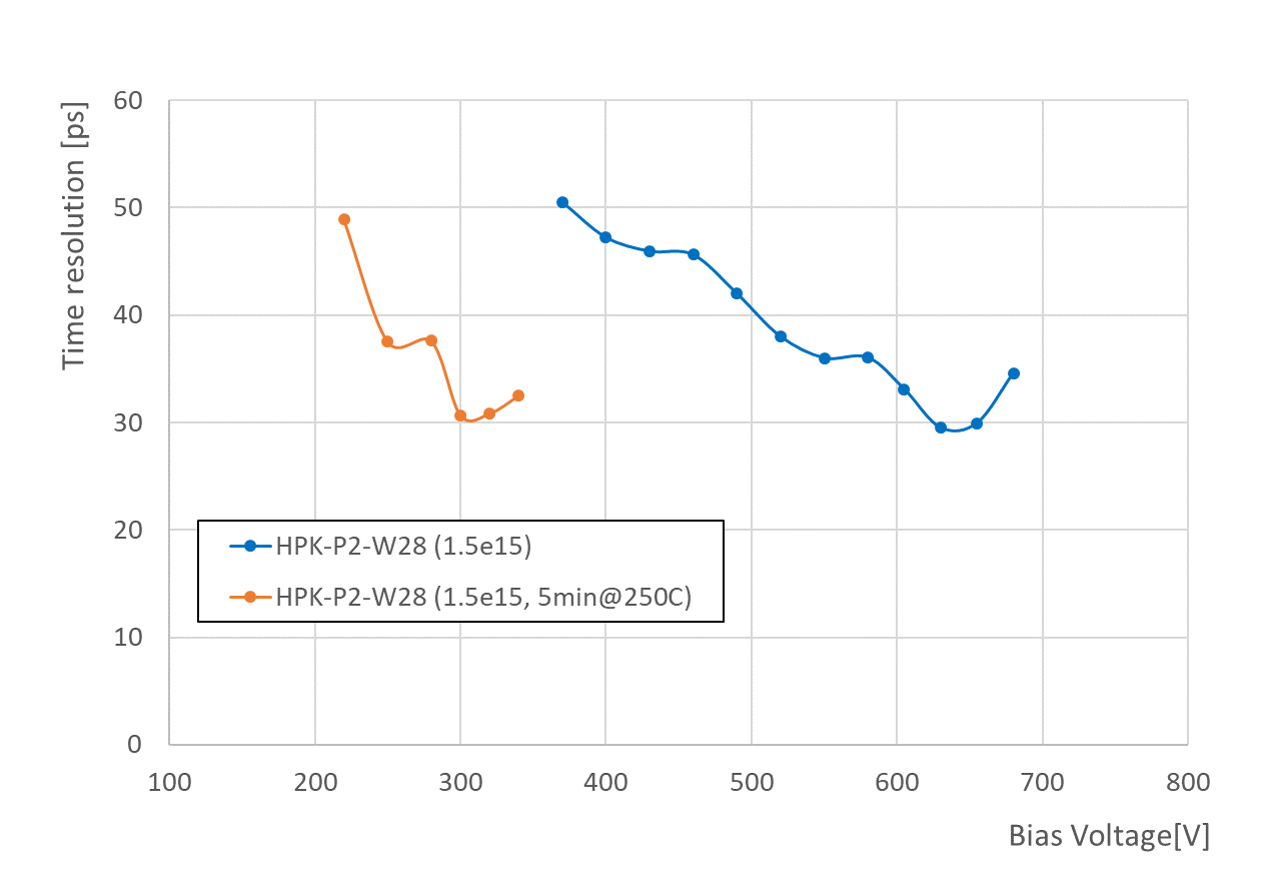} \\
 a) & b) \\  
 \end{tabular}
 \caption{\label{Timing-5min} Measurements with $^{90}$Sr source. Figure a) show collected charge vs. bias voltage before and after annealing for 5 minutes at 250$^\circ$C with device irradiated to 1.5$\cdot10^{15}$ n$_\mathrm{eq}$/cm$^2$. Figure b) timing resolution as a function of bias voltage and figure c) show leakage current measured on timing setup at temperature of -30$^\circ$C.}
\end{figure}
Large changes of full depletion and gain layer depletion voltage can be seen: before annealing full depletion voltage was below 100 V while after annealing it is above 350 V (see Fig. \ref{CVIV-5min}a), the maximum at which measurements were taken. More important for LGAD is $V_{\mathrm{gl}}$, determined from the location of the first knee of C-V graph in Fig. \ref{CVIV-5min}b), which increases after annealing. This drop of capacitance is the consequence of gain layer depletion and $V_{\mathrm{gl}}$ is the bias voltage at which it occurs. $V_{\mathrm{gl}}$ is defined as the intersection of lines fitted to straight parts on each side of the knee as shown in  Fig. \ref{CVIV-5min}b).
Figure \ref{CVIV-5min}c) shows the comparison of I-V measurements before and after annealing. It can be seen that the leakage current is lower by an order of magnitude after annealing.

Figure \ref{Timing-5min} shows the results of measurements with $^{90}$Sr and significant improvement of charge collection and timing resolution after annealing can be seen. The voltage at which 10 fC charge is collected V$_\mathrm{10fC}$ drops from close to 600 V to below 300 V and the same timing resolution is reached at much lower bias voltage after annealing than before.

The improvement of performance can be explained by the increase of space charge concentration caused by the annealing. In the depleted gain layer the increase of space charge concentration increases the electric field which leads to larger charge multiplication. The increase of space charge concentration in the gain layer can be attributed to re-activation of boron dopants by high temperature treatment which causes a change of their state from interstitial to substitutional (B$_I \rightarrow$ B$_S$). The increase of space charge concentration in the gain layer needed for the observed change of $V_{\mathrm{gl}}$ is an order of magnitude higher than the increase of space charge concentration due to ``standard'' reverse annealing of deep (radiation induced) defects \cite{ROSE}. The latter is the dominant process which increases space charge concentration in the bulk of the device causing increase of full depletion voltage $V_{\mathrm{fd}}$. It is somewhat less straightforward to see that increase of $V_{\mathrm{fd}}$ may also lead to better performance at lower bias voltage. Namely, to establish sufficient electric field strength in the gain layer, lower bias voltage needs to be applied because voltage drop over the fully depleted bulk is smaller if space concentration is higher and so the electric field profile in the bulk is steeper - see \cite{lgad_annealing}. The improvement of charge collection and timing properties after annealing for 5 minutes at 250$^\circ$C is predominantly attributed to this effect because the change of $V_{\mathrm{gl}}$ is too small.

\subsection{Annealing at higher temperatures}

\begin{figure}[!hbt]
\centering
\begin{tabular}{c c} 
 \includegraphics[width=0.5\textwidth]{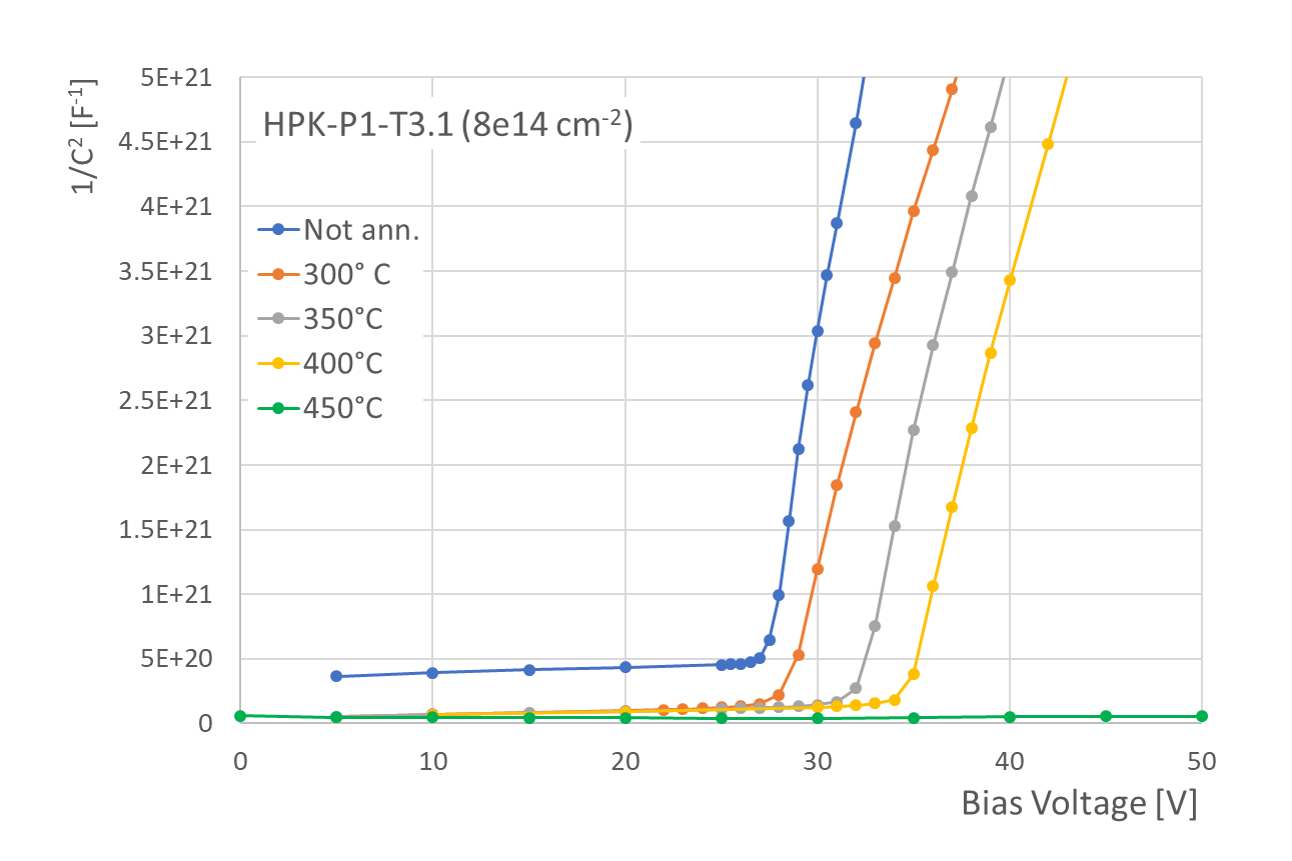} &  \includegraphics[width=0.5\textwidth]{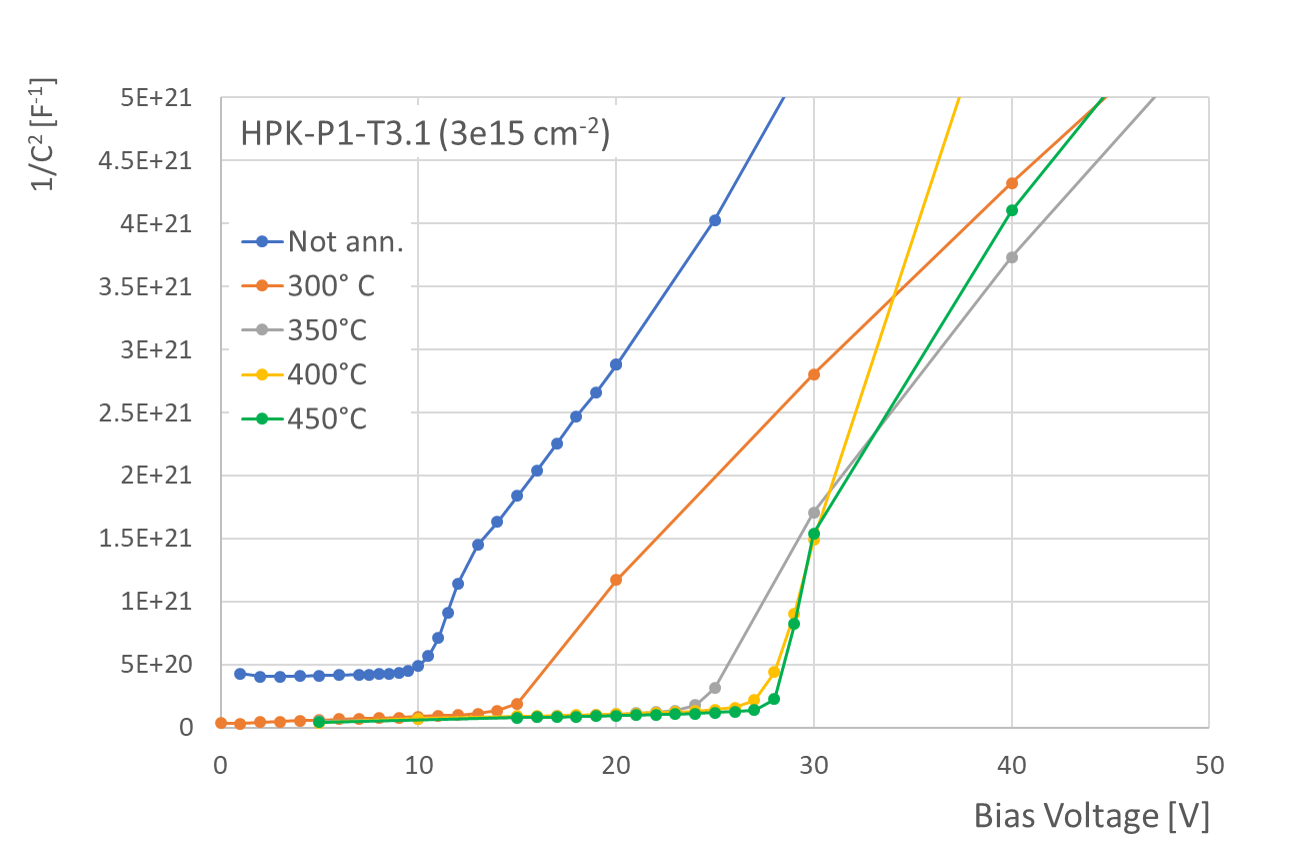} \\
 a) & b) \\
 \includegraphics[width=0.5\textwidth]{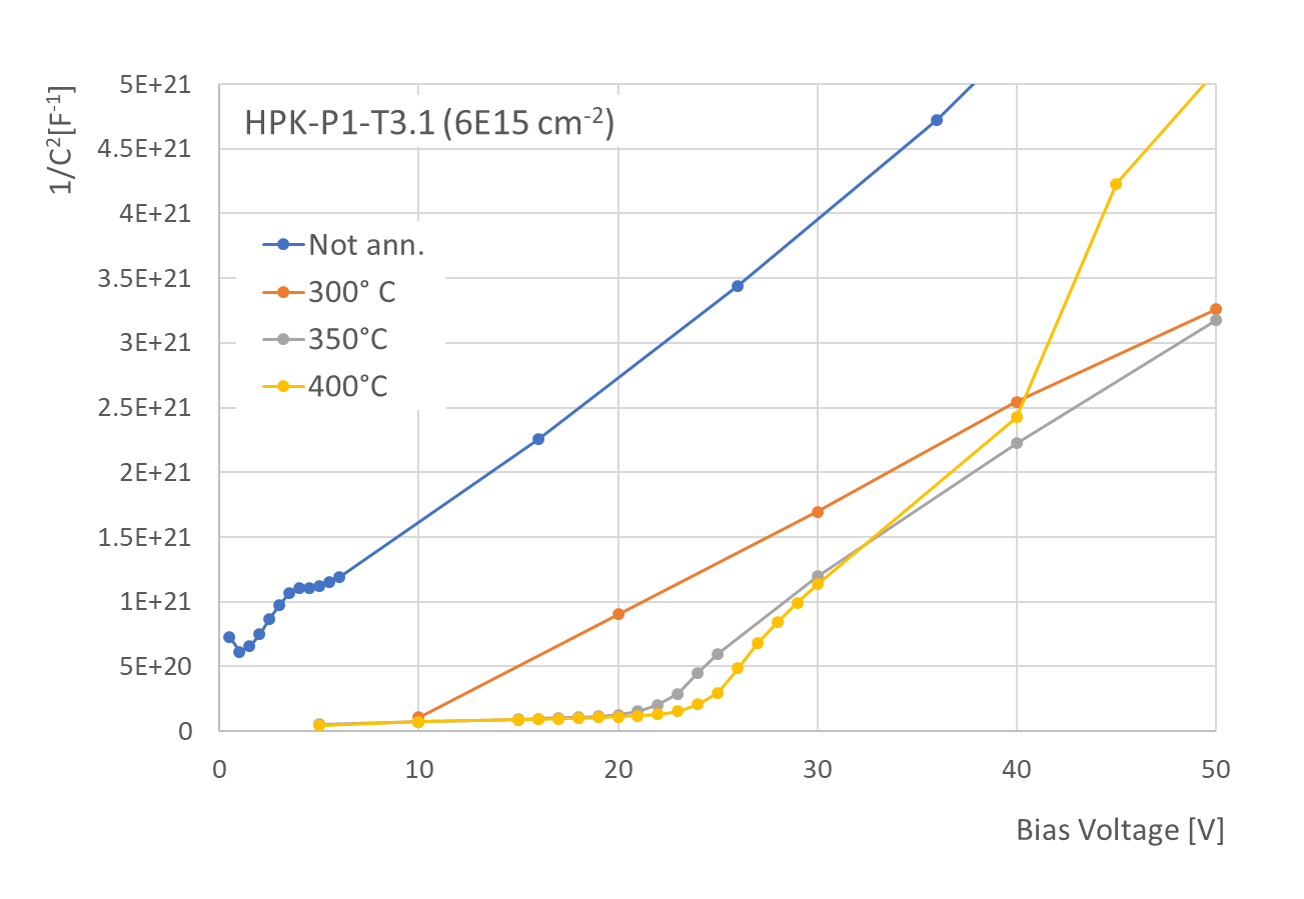} &  \includegraphics[width=0.5\textwidth]{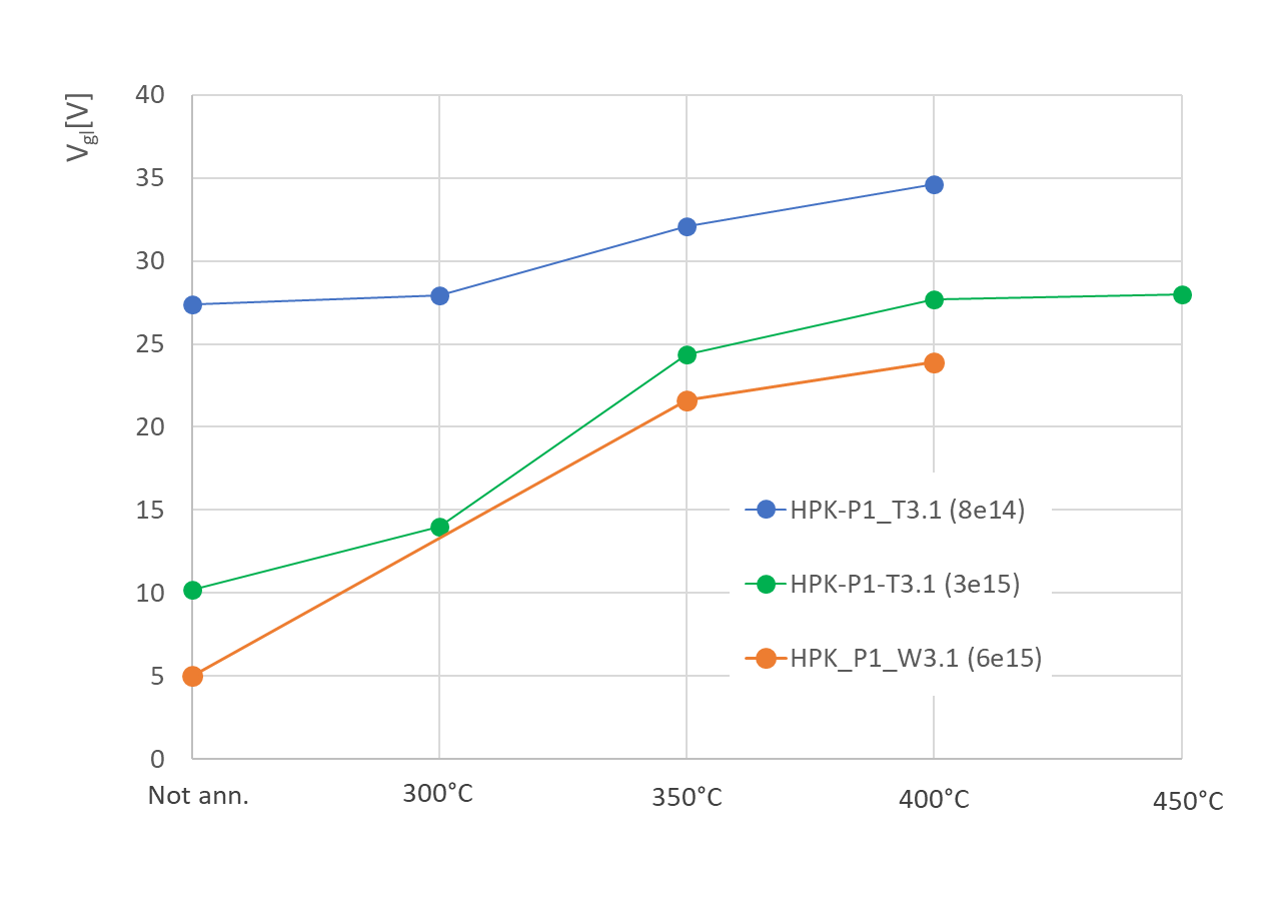} \\
 c) & d) 
 \end{tabular}
 \caption{\label{CV-isochronal} Figures a), b) and c) show CV measurements after 30 minute annealing steps: step 0: before annealing, step 1: 300$^\circ$C, step 2: 350$^\circ$C, step 3: 400$^\circ$C and step 4: 450$^\circ$C for LGADs irradiated to different fluences. Figure d) shows V$_{\mathrm{gl}}$ estimated from CV measurements shown in a), b) and c) as a function of step number.}
\end{figure}

For studies at higher temperatures, irradiated LGADs were heated in argon atmosphere with temperature controlled with accuracy of few degrees. 
In this study isochronal steps of 30 minutes were taken at 300$^\circ$C, 350$^\circ$C, 400$^\circ$C and 450$^\circ$C with three devices of type HPK-P1-T3.1 irradiated to fluences listed in table \ref{tab1}. After each annealing step C-V and I-V measurements were performed. 
Figures \ref{CV-isochronal}a), \ref{CV-isochronal}b) and \ref{CV-isochronal}c) show C-V curves for the three devices after each annealing step. V$_{\mathrm{gl}}$ increase with each annealing step can clearly be seen and the effect is the largest in the LGAD irradiated to the highest fluence. Measurement at low bias voltages after step 1 for the sample irradiated to 6e15 n/cm$^2$ (Fig. \ref{CV-isochronal}c) were too noisy and V$_{\mathrm{gl}}$ could not be extracted but it can be seen it was not higher than 10 V. After step 4 (450$^\circ$C) the aluminium contact pads were damaged due to the high temperature and good C-V measurement could only be made with the sample irradiated to 3e15 n/cm$^2$ (Fig. \ref{CV-isochronal}b).

Figure \ref{CV-isochronal}d) shows the evolution of V$_{\mathrm{gl}}$ estimated from C-V curves. The largest increase of V$_{\mathrm{gl}}$ is observed after annealing for 30 minutes at 350$^\circ$C. 

\begin{figure}[!hbt]
\centering
\begin{tabular}{c c} 
 \includegraphics[width=0.5\textwidth]{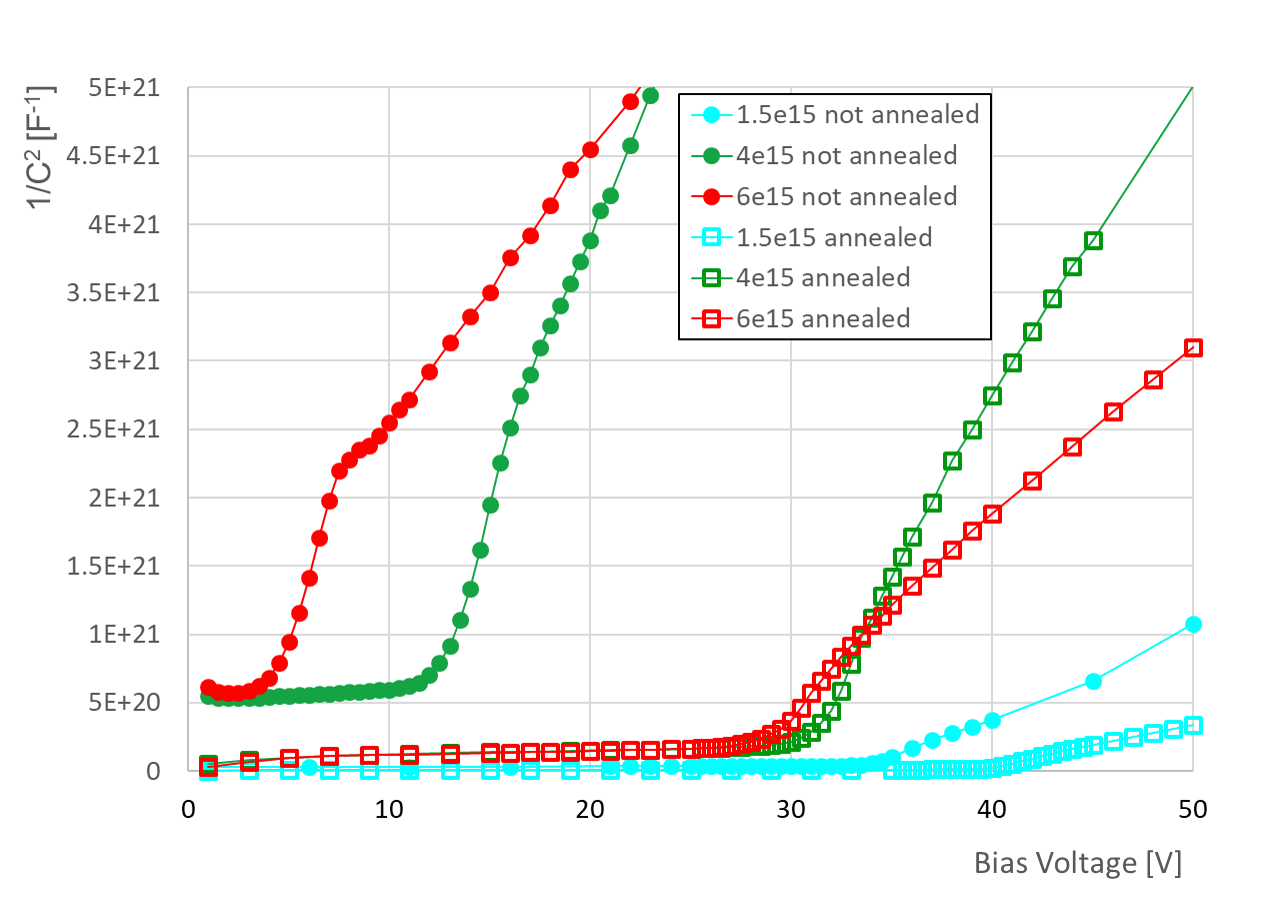} &  \includegraphics[width=0.5\textwidth]{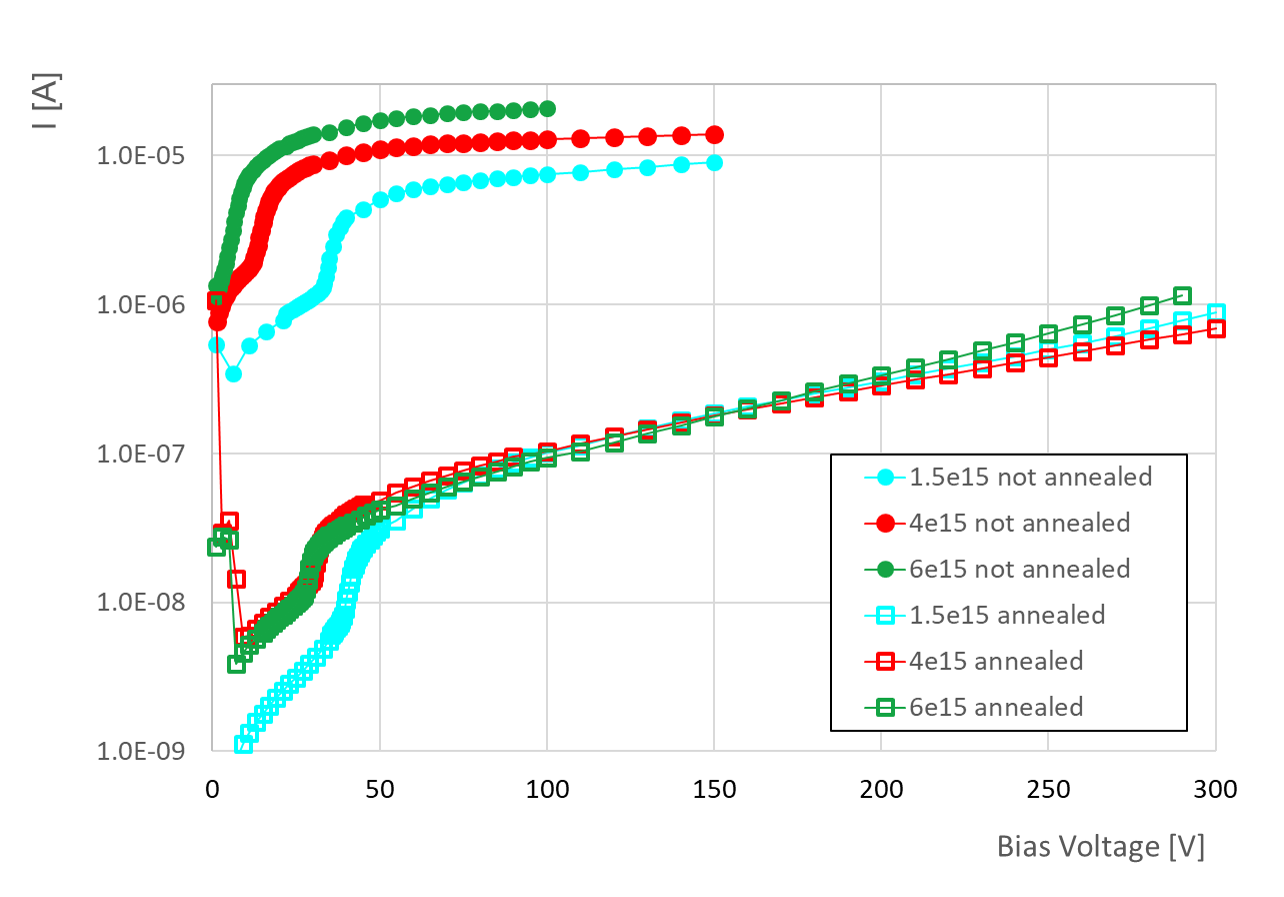} \\
 a) & b) \\
  \end{tabular}
\begin{tabular}{c} 
  \includegraphics[width=0.5\textwidth]{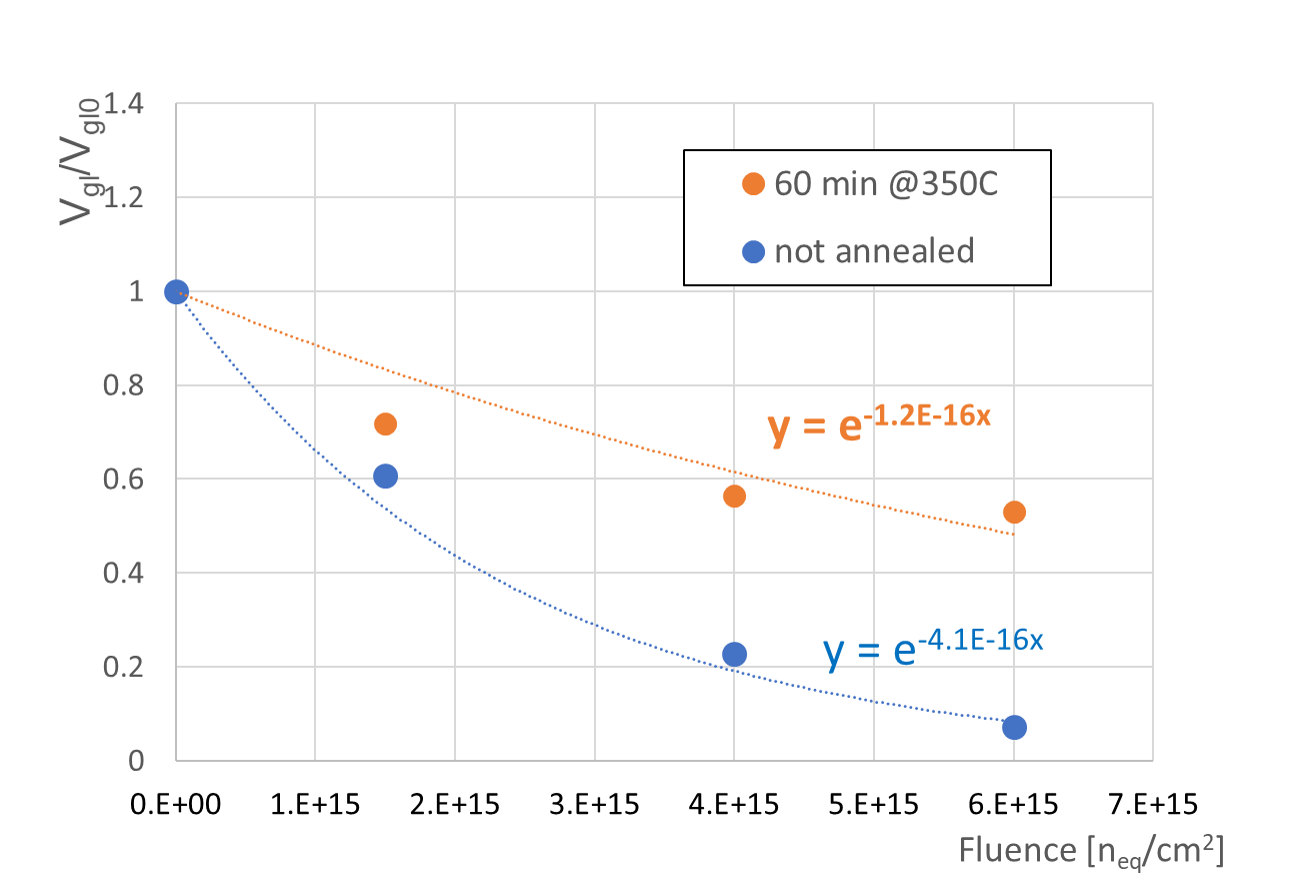}   \\
  c)  \\
 \end{tabular}
 \caption{\label{CV-IV-60minutes} a) C-V measurement with three LGADs of type HPK-P1-T3.2 before and after annealing for 60 minutes at 350$^\circ$C. Figure b) shows I-V measurements for these devices. Figure c) shows $V_{\mathrm{gl}}$ extracted from Fig a) relative to the pre irradiation value $V_{\mathrm{gl0}}$ as a function of fluence. Exponential functions are fitted to measured points.}
\end{figure}

Following these results another experiment was performed in which three LGADs of type HPK-P1-T3.2 irradiated to different fluences (see table \ref{tab1}) were annealed in a single annealing step for 60 minutes at 350$^\circ$C.  Figure \ref{CV-IV-60minutes} shows C-V and I-V measurements with these three samples.  Increase of V$_{\mathrm{gl}}$ can clearly be seen as well as an almost 2 orders of magnitude large drop of reverse current. Figure \ref{CV-IV-60minutes}c) shows the change of $V_{\mathrm{gl}}$ relative to the pre irradiation value $V_{\mathrm{gl0}}$ as a function of fluence. Exponential functions were fitted to annealed and non-annealed points. The value of $V_{\mathrm{gl}}$ is proportional to the space charge concentration in the gain layer so the coefficient in the exponent represents an effective acceptor removal constant. It can be seen that annealing for 60 minutes at 350$^\circ$C reduces this constant by a factor of $\sim$ 4.  

\begin{figure}[!hbt]
\centering
\begin{tabular}{c c} 
 \includegraphics[width=0.5\textwidth]{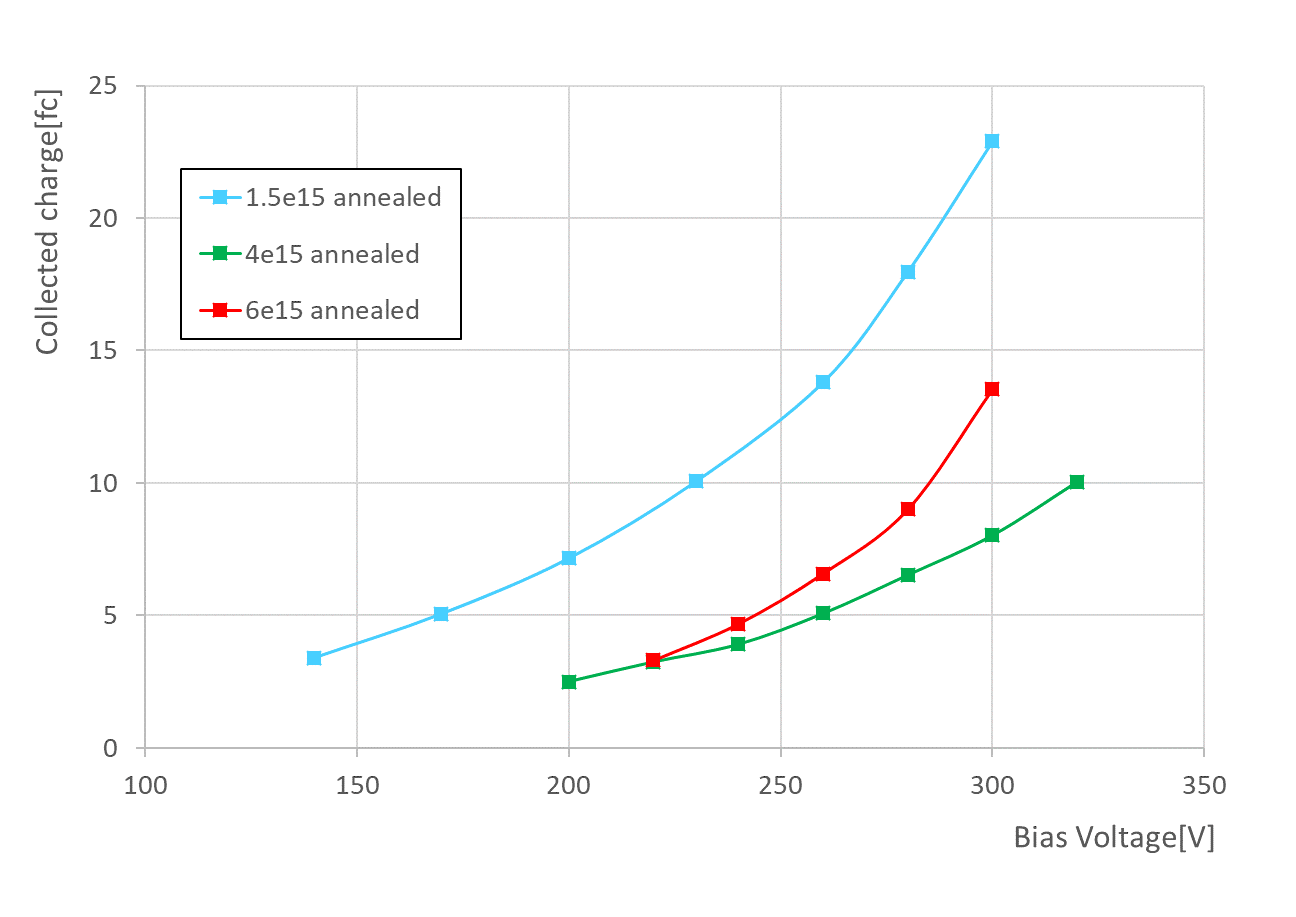} &  \includegraphics[width=0.5\textwidth]{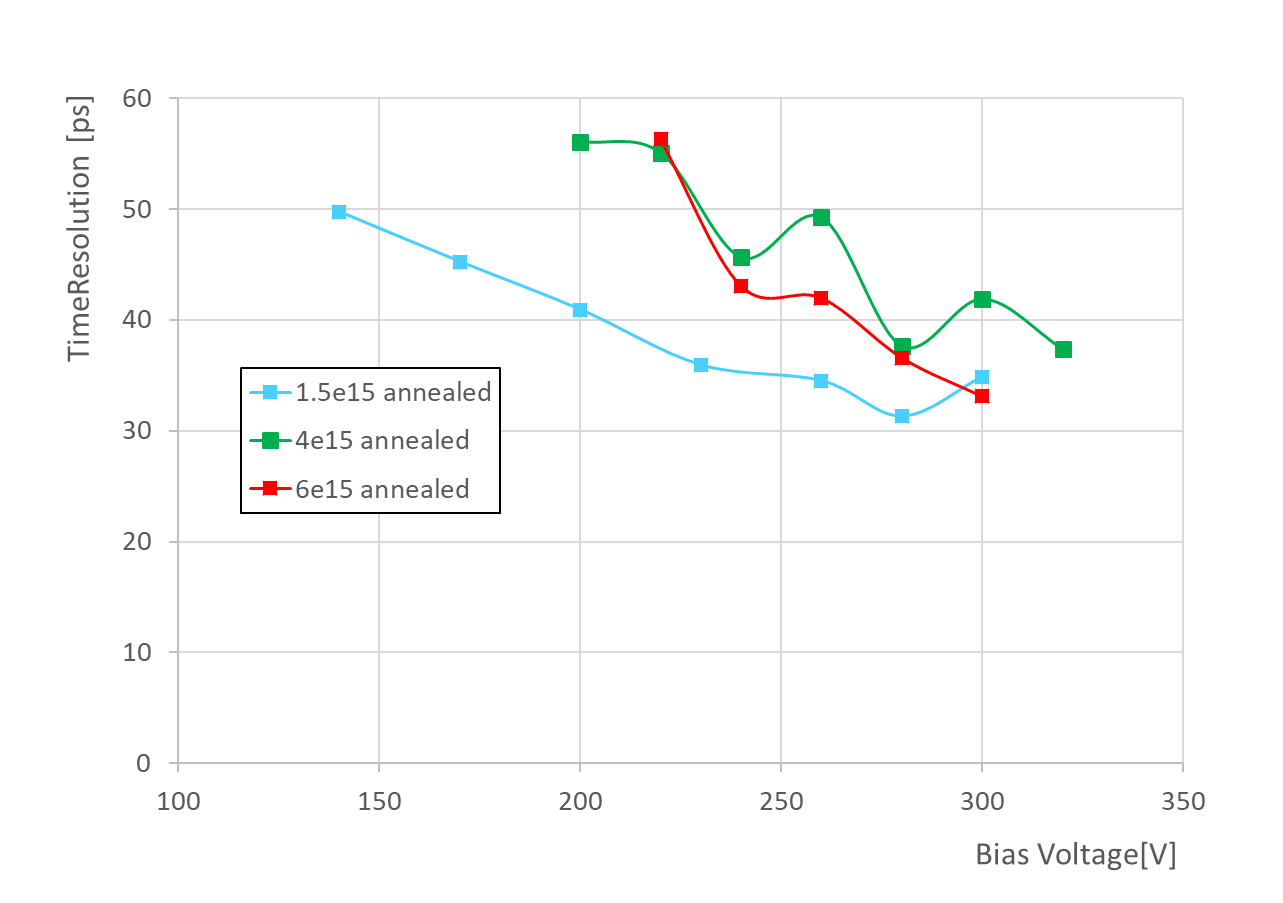} \\
 a) & b) 
 \end{tabular}
 \caption{\label{Timming-60min} a) Collected charge vs. bias voltage measured with $^{90}$Sr source for irradiated HPK-P1-T3.2 devices after annealing for 60 minutes at 350$^\circ$C. In figure b) time resolution as a function of bias voltage is shown.}  
\end{figure}

Measurements of these devices were made also with $^{90}$Sr source. Collected charge and time resolution as a function of bias voltage are shown in figure \ref{Timming-60min}. As can be seen, annealing for 60 minutes at 350$^\circ$C greatly improves charge collection and time resolution, and restores the performance of heavily irradiated LGADs to almost the pre-irradiation level. Before high temperature annealing, gain operation would not be possible below 350 V after irradiation to such high fluences \cite{lgad_annealing, ucsc}. 

\begin{figure}[!hbt]
\centering
\begin{tabular}{c c} 
 \includegraphics[width=0.5\textwidth]{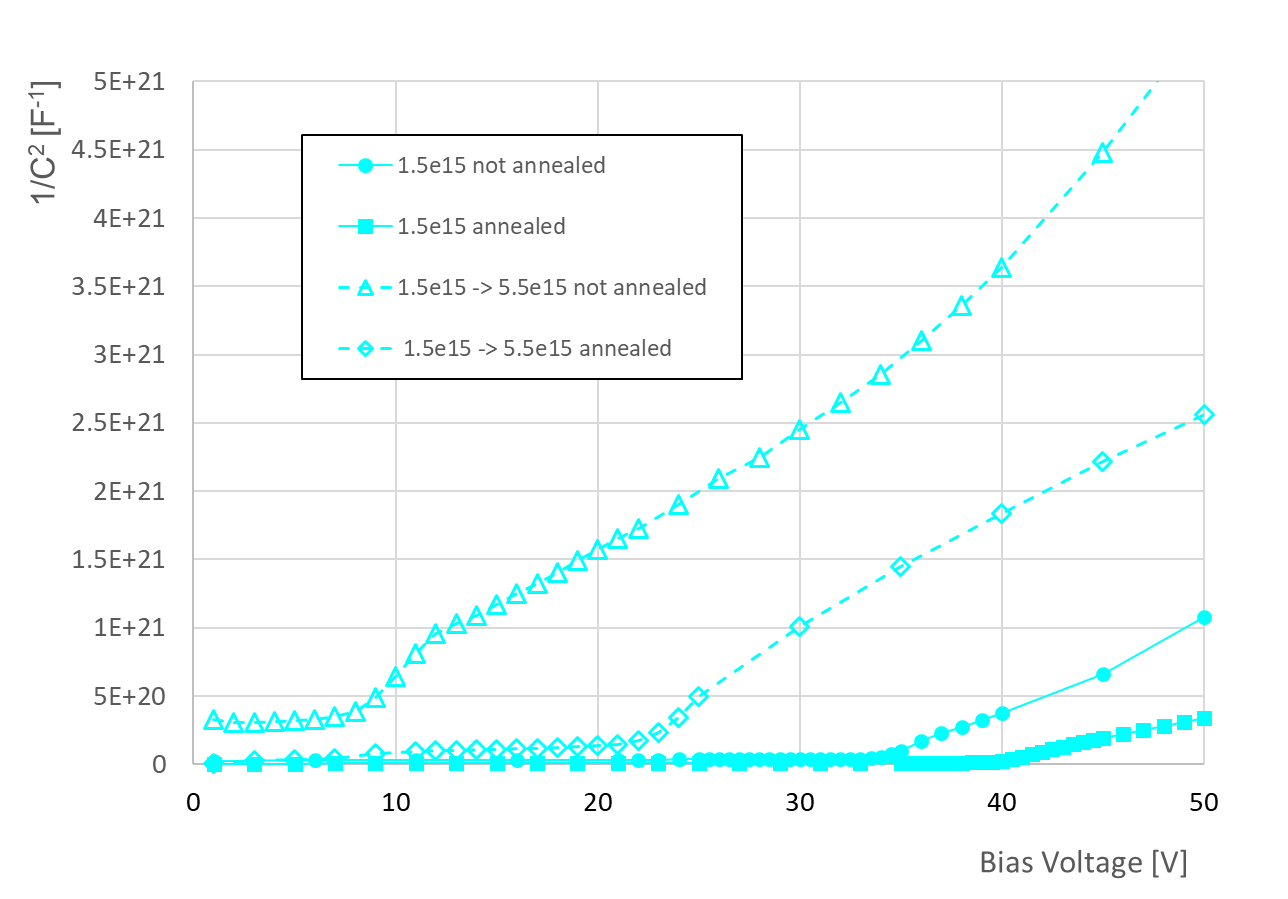} &  \includegraphics[width=0.5\textwidth]{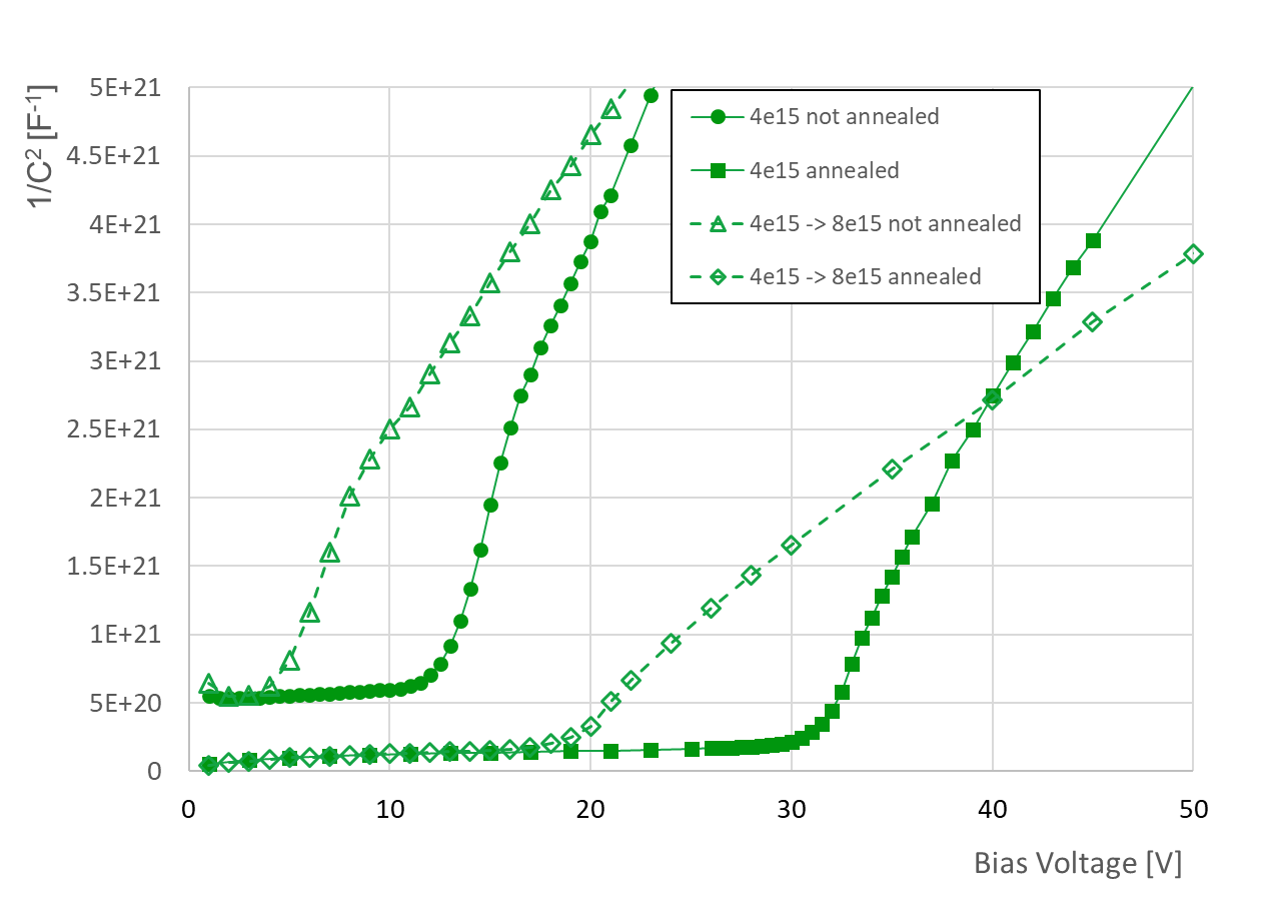} \\
 a) & b) \\
  \includegraphics[width=0.5\textwidth]{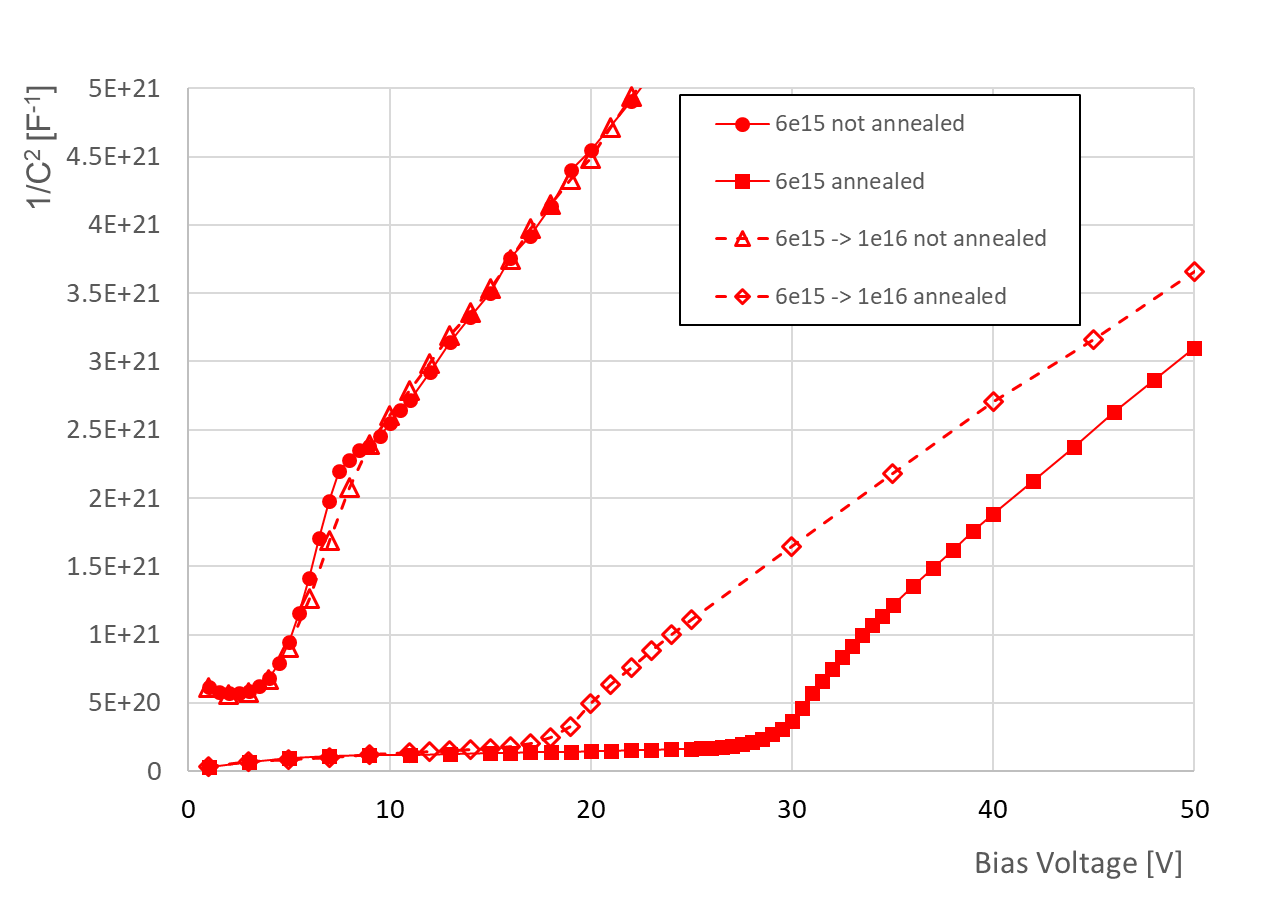}  \\
 c) & \\
 \end{tabular}
 \caption{\label{CV-60min-step1} CV measurements for HPK-P1-3.2 devices measured after two irradiation and annealing steps. Full lines are measurements after first irradiation and dashed lines show measurements after another irradiation with 4e15 n/cm$^2$. Figure a) shows 1.5e15 n/cm$^2$ and 5.5e15 n/cm$^2$, figure b) 4e15 n/cm$^2$ and 8e15 n/cm$^2$ and c) 6e15 n/cm$^2$ and 1e16 n/cm$^2$.}  
\end{figure}

Good results encouraged investigations of performance of LGADs at even higher fluences. For this purpose the irradiated and annealed HPK-P1-T3.2 devices were irradiated with additional 4e15 n/cm$^2$ neutrons in the reactor in Ljubljana and again annealed for 60 minutes at 350$^\circ$C. However, this time annealing was performed with different equipment in standard atmosphere. After 60 minutes at 350$^\circ$C the furnace was switched off, however temperature stayed high (above 200$^\circ$C) for another few hours. Therefore, this annealing step is not directly comparable to previous ones. C-V curves measured after additional irradiation before and after annealing are shown in Figure \ref{CV-60min-step1}.  Additional irradiation again reduced V$_{\mathrm{gl}}$ which was subsequently significantly increased by annealing. In \ref{CV-60min-step1}c) it can be seen that after annealing $V_{\mathrm{gl}} \sim$ 30 V was measured for LGAD irradiated to 1e16 n/cm$^2$. This is a very high value for such a high fluence and indicates that gain in LGADs might be restored even after such heavy irradiation. Unfortunately, measurements of timing performance could not be made because the noise was of too high, with the origin not clear to us, so further studies are needed to clarify this.

\section{Conclusions}

In this work the effect of annealing on irradiated LGADs at very high temperatures in the range from 250$^\circ$C to 450$^\circ$C was investigated. 

Annealing for 5 minutes at 250$^\circ$C slightly increased the gain layer depletion voltage $V_{\mathrm{gl}}$ and had a large impact on full depletion $V_{\mathrm{fd}}$. Both can be attributed to the increased space charge concentration due to reactivation of boron dopants and reverse annealing of deep defects. Annealing significantly improved performance of devices because it allowed the operation voltage to stay well below the voltage of single event burnout and to achieve sufficient gain necessary for good timing resolution. The improved performance was mainly the consequence of increased effective space charge concentration in the bulk which reduced voltage drop over the depleted bulk resulting in lower bias voltage needed for sufficient electric field strength in the gain layer.

Isochronal annealing study with 30 minutes intervals at temperatures between 300$^\circ$C and 450$^\circ$C showed that in this temperature range annealing increases $V_{\mathrm{gl}}$ and that the increase is most significant up to 350$^\circ$C. Following these findings another set of irradiated devices were annealed for 60 minutes at 350$^\circ$C which greatly increased $V_{\mathrm{gl}}$. Dependence of $V_{\mathrm{gl}}$
on fluence (Fig. \ref{CV-IV-60minutes}c) can be approximated with an exponential function and it shows that annealing reduces the effective acceptor removal constant by a factor of $\sim$ 4. 

After annealing the timing properties of the devices were characterized with $^{90}$Sr and a timing resolution close to 30 ps could be achieved at bias of 300 V with LGAD irradiated with 6e15 n/cm$^2$. This is very good performance, comparable to that before irradiation. Initial results of high temperature annealing at fluences up to 1e16 n/cm$^2$ are also encouraging and it can be hoped that their operating range can be extended up to such high radiation levels.   

Annealing of irradiated silicon detectors at elevated temperatures to deal with radiation damage was proposed a long time ago in \cite{DRIVE}. Positive effects on full depletion voltage were seen after annealing at 450$^\circ$C and the effect was attributed to generation of thermal donors which compensated radiation induced negative space charge. Heating silicon detectors to such high temperatures in real experiment is a difficult technical problem but several possibilities were proposed in \cite{DRIVE} such as localised laser heating or a pre-built-in external heating circuit.

It should be noted that LGADs are better suited for application of annealing because lower temperatures are needed to improve their performance. The reason is that unlike in standard detectors, in LGADs the increase of negative space charge concentration (i.e. by reverse annealing) in the gain layer and in the bulk are beneficial. Another important aspect of LGADs is that improvement might be achieved by heating only the very thin gain layer while keeping other sensitive parts of the detector system at lower temperatures. An example of such approach can be found in \cite{tempdiode} where radiation damage in temperature sensing diodes was mitigated by heating with tungsten heater integrated in the device.

\section{Acknowledgments}

The authors would like to thank the crew at the TRIGA reactor in Ljubljana for help with irradiation of detectors.
Part of this work was performed in the framework of the CERN-RD50 collaboration. The authors acknowledge the financial
support from the Slovenian Research Agency (research core funding No. P1-0135 and project No. J1-1699).

\end{document}